\documentclass[useAMS,usenatbib]{mn2e}

\usepackage{graphicx}
\usepackage{rotating}
\usepackage{longtable}
\usepackage{lscape}

\def\Msun{M$_\odot$}
\def\Mbh{${\cal M}_{\rm BH}$}
\def\Lhost{$L_{\rm host}$}
\def\Mhost{${\cal M}_{\rm host}$}

\def\hi{H\,{\sc i}}
\def\Civ{C\,{\sc iv}}
\def\Mgii{Mg\,{\sc ii}}

\def\Oiii{[O\,{\sc iii}]}
\def\Hb{H$\beta$}
\def\lsim{\mathrel{\rlap{\lower 3pt \hbox{$\sim$}} \raise 2.0pt \hbox{$<$}}}
\def\gsim{\mathrel{\rlap{\lower 3pt \hbox{$\sim$}} \raise 2.0pt \hbox{$>$}}}

\title[The \Mbh--\Mhost{} relation through Cosmic Time]{
The quasar \Mbh--\Mhost{} relation through Cosmic Time\\ II -- Evidence for evolution from $z=3$ to the present age}
\author[Decarli et al.]{
R. Decarli$^{1,2}$\thanks{E-mail: decarli@mpia-hd.mpg.de},
R. Falomo$^{3}$,
A. Treves$^{1,4}$,
M. Labita$^{1}$,
J.K. Kotilainen$^{5}$,
and R. Scarpa$^{6}$
\\
$^{1}$ Universit\`{a} degli Studi dell'Insubria, via Valleggio 11,
             I-22100 Como, Italy\\
$^{2}$ Max Planck Institute f\"ur Astronomie, K\"onigsthul 17, D-69117 Heidelberg, Germany\\
$^{3}$ INAF - Osservatorio Astronomico di Padova, Vicolo dell'Osservatorio 5,
             I-35122, Padova, Italy\\
$^{4}$ Istituto Naz. Fis. Nucleare, Piazza della Scienza 3, I-20126 Milano, Italy\\
$^{5}$ Tuorla Observatory, Department of Physics and Astronomy,
University of Turku, V\"ais\"al\"antie 20, FI-21500 Piikki\"o, Finland\\
$^{6}$ Instituto de Astrofisica de Canarias, Via Lactea, s/n E-38205
La Laguna (Tenerife), Spain
         }
\begin{document}

\date{ }

\pagerange{\pageref{firstpage}--\pageref{lastpage}} \pubyear{2009}

\maketitle

\label{firstpage}

\begin{abstract}
We study the dependence of the \Mbh{}--\Mhost{} relation on the
redshift up to $z=3$ for a sample of 96 quasars the host galaxy
luminosities of which are known. Black hole masses were estimated
assuming virial equilibrium in the broad line regions (Paper I),
while the host galaxy masses were inferred from their luminosities.
With this data we are able to pin down the redshift dependence of
the \Mbh{}--\Mhost{} relation along 85 per cent of the Universe age.
We show that, in the sampled redshift range, the \Mbh{}--\Lhost{}
relation remains nearly unchanged. Once we take into account the
aging of the stellar population, we find that the \Mbh{}/\Mhost{}
ratio ($\Gamma$) increases by a factor $\sim 7$ from $z=0$ to $z=3$.
We show that $\Gamma$ evolves with $z$ regardless of the radio
loudness and of the quasar luminosity. We propose that most massive
black holes, living their quasar phase at high-redshift, become
extremely rare objects in host galaxies of similar mass in the Local
Universe.
\end{abstract}


\section{Introduction}

Many pieces of evidence suggest that supermassive black holes (BHs)
and their host galaxies share a joint evolution throughout Cosmic
Time. In particular: {\it i-} the evolution of the quasar luminosity
function \citep{dunlop90,fontanot07,croom09} closely matches the
trend of the star formation density through Cosmic ages
\citep{madau98}; {\it ii-} massive black holes are found in
virtually all massive galaxies \citep{kormendy95,decarli07}; {\it
iii-} their mass (\Mbh{}) is tightly correlated with the large scale
properties (stellar velocity dispersion, $\sigma_*$; luminosity,
\Lhost{}; mass, \Mhost{}) of their host galaxies \citep[see][for
recent reviews on this topic]{ferrarese06,gultekin09}.

When and how these relations set in, and which are the physical
processes responsible of their onset are still open questions,
despite the large efforts perfused both from a theoretical
\citep[e.g.,][]{silk98,king05,wyithe06,robertson06,hopkins07,malbon07}
and an observational point of view
\citep{mclure06,peng06a,peng06b,salviander07,woo08}.

Probing BH--host galaxy relations at high redshift is extremely
challenging. Direct measurements of \Mbh{} from gaseous or stellar
dynamics require observations capable to resolve the BH sphere of
influence, which are feasible only for a limited number of nearby
galaxies. The only way to measure \Mbh{} in distant ($>$ few tens of
Mpc) galaxies is to focus on Type-1 AGN, where \Mbh{} can be
inferred from the width of emission lines Doppler-broadened by the
BH potential well and from the AGN continuum luminosity \citep[see,
e.g.,][]{vestergaard02}, assuming the virial equilibrium. Quasars
represent therefore the best tool to probe \Mbh{} at high redshift,
thanks to their huge luminosity. Indeed, large-field spectroscopic
surveys allowed the estimate of \Mbh{} in several thousands of
objects \citep[see, for instance,][]{shen08,labita09a,labita09b}. On
the other hand, the AGN light in quasars outshines the emission from
the host galaxies, making their detection challenging. Only recently
limitations due to intrinsic (e.g. the Nucleus-to-Host galaxy
luminosity ratio, N/H) and extrinsic (e.g., the angular size of the
host with respect to the angular resolution of the observations)
effects could be overcome. Optical images from the Hubble Space
Telescope (HST) and NIR observations both from space and
ground-based telescopes could resolve $\sim300$ quasar host galaxies
up to $z\sim3$ \citep[see, e.g.,][and references
therein]{kotilainen09}. Preliminary studies suggest that, for a
given galaxy mass, black holes in high-$z$ AGN are more massive than
their low-$z$ counter-parts
\citep[e.g.,][]{mclure06,peng06a,peng06b}.

A number of limitations potentially affect the studies of the
\Mbh{}--\Mhost{} relation through Cosmic Time:
{\bf 1)} All the works to date use different \Mbh{} proxies as a function
of redshift (i.e., based on different broad emission lines: usually \Hb{} at
$z\lsim0.5$, \Mgii{} for $0.5 \lsim z \lsim 2$ and \Civ{} at $z\gsim1.6$).
{\bf 2)} Selection biases related to luminosity or flux limits, to the
sampled N/H, to the steepness of the bright
end of the galaxy and quasar luminosity functions may hinder the study of
the evolution of the \Mbh{}--\Mhost{} relation \citep[see, for instance,][]{lauer07}.
{\bf 3)} As the properties of quasar host galaxies are directly observed only
in a limited number of objects, poor statistics usually affect the available
datasets.

This study represents a significant effort in overcoming all these limitations:
Thanks to UV and optical spectra of low-redshift quasars
\citep{labita06,decarli08} and to optical spectra of mid- and high-redshift quasars
\citep[][hereafter Paper I]{paperI} we can probe \Mbh{} using both
high and low ionization lines in a wide range of redshifts, for the largest dataset
adopted so far in this kind of studies.

In Paper I we presented the sample and we inferred BH masses.
Here we describe the data sources for the host galaxy luminosities,
we infer \Mhost{} and we address the evolution of the
\Mbh{}--\Mhost{} relation.

Throughout the paper, we adopt a concordance cosmology
with $H_0=70$ km/s/Mpc, $\Omega_m=0.3$, $\Omega_\Lambda=0.7$.
We converted the results of other authors to this cosmology when
adopting their relations and data.

\section[]{\Lhost{} and \Mhost{} in distant quasars}\label{sec_host}
\begin{table*}
\begin{center}
\caption{The sample in this study. (1) Quasar name. (2) Catalogue redshift.
(3) Host galaxy $R$-band absolute magnitude (not corrected for stellar aging). (4)
Host galaxy stellar mass. (5) Reference for the host galaxy luminosity: 1- low-$z$ HST-based observation
(see Paper I); 2- \citet{falomo04}; 3- \citet{kukula01}; 4- \citet{ridgway01};
5- \citet{falomo05}; 6- \citet{kotilainen09}; 7- \citet{kotilainen07};
8- \citet{hyvonen07a}; 9- \citet{kotilainen98}; 10- \citet{kotilainen00}; 11- \citet{decarli09a}.
}\label{tab_sample}
\begin{tabular}{ccccccccccccc}
\hline
                  &       &\vline&\multicolumn{3}{c}{Host galaxy}& \vline{} \vline &                      &      &\vline&\multicolumn{3}{c}{Host galaxy}\\
 Quasar Name          & $z$   &\vline& $M_R$  & log \Mhost{} & Ref   & \vline{} \vline &  Quasar Name         & $z$  &\vline& $M_R$ & log \Mhost{} & Ref    \\
                  &       &\vline& [mag]  & [\Msun]      & $M_R$ & \vline{} \vline &                  &      &\vline& [mag] & [\Msun]      & $M_R$  \\
 (1)                  & (2)   &\vline& (3)    & (4           & (5)   & \vline{} \vline &  (1)                 & (2)  &\vline& (3)   & (4)      & (5)    \\
\hline
          PKS0000-177 & 1.465 &\vline&  -24.5 & 12.0 &  2  & \vline{} \vline &       1116+215 & 0.177 &\vline&  -22.2 &  11.5 &  1  \\
           Q0040-3731 & 1.780 &\vline&  -22.8 & 11.3 &  2  & \vline{} \vline &       1150+497 & 0.334 &\vline&  -23.4 &  11.9 &  1  \\
              SGP2:36 & 1.756 &\vline&  -23.7 & 11.7 &  3  & \vline{} \vline &       1202+281 & 0.165 &\vline&  -21.4 &  11.2 &  1  \\
              SGP5:46 & 0.955 &\vline&  -22.4 & 11.4 &  3  & \vline{} \vline &       1208+322 & 0.388 &\vline&  -23.5 &  12.0 &  1  \\
             0054+144 & 0.171 &\vline&  -22.9 & 11.8 &  1  & \vline{} \vline &       1216+069 & 0.331 &\vline&  -22.1 &  11.4 &  1  \\
              SGP4:39 & 1.716 &\vline&  -21.7 & 10.9 &  3  & \vline{} \vline &        MRK0205 & 0.071 &\vline&  -23.3 &  12.0 &  1  \\
          PKS0100-270 & 1.597 &\vline&  -23.4 & 11.6 &  2  & \vline{} \vline &       1222+125 & 0.415 &\vline&  -23.7 &  12.0 &  1  \\
        LBQS0100+0205 & 0.393 &\vline&  -22.3 & 11.5 &  1  & \vline{} \vline &      3C273 & 0.158 &\vline&  -22.8 &  11.8 &  1  \\
             0110+297 & 0.363 &\vline&  -23.6 & 12.0 &  1  & \vline{} \vline &       1230+097 & 0.415 &\vline&  -24.0 &  12.2 &  1  \\
          PKS0113-283 & 2.555 &\vline&  -24.7 & 11.8 &  5  & \vline{} \vline &   Z124029-0010 & 2.030 &\vline&  -25.5 &  12.3 &  6  \\
             0119-370 & 1.320 &\vline&  -23.7 & 11.8 &  2  & \vline{} \vline &     PG1302-102 & 0.286 &\vline&  -23.1 &  11.8 &  1  \\
             0133+207 & 0.425 &\vline&  -22.8 & 11.7 &  1  & \vline{} \vline &       1307+085 & 0.155 &\vline&  -21.1 &  11.1 &  1  \\
                 3C48 & 0.367 &\vline&  -25.6 & 12.8 &  1  & \vline{} \vline &       1309+355 & 0.184 &\vline&  -22.6 &  11.7 &  1  \\
         HB890137+012 & 0.260 &\vline&  -23.5 & 12.0 &  1  & \vline{} \vline &       1402+436 & 0.320 &\vline&  -23.3 &  11.9 &  1  \\
            0152-4055 & 1.650 &\vline&  -23.4 & 11.6 &  2  & \vline{} \vline &     PG1416-129 & 0.129 &\vline&  -21.7 &  11.3 &  1  \\
          PKS0155-495 & 1.298 &\vline&  -24.4 & 12.0 &  2  & \vline{} \vline &       1425+267 & 0.366 &\vline&  -24.3 &  12.3 &  1  \\
           PKS0159-11 & 0.669 &\vline&  -22.3 & 11.4 & 10  & \vline{} \vline &   Z143220-0215 & 2.476 &\vline&  -23.2 &  11.2 &  6  \\
           B0204+2916 & 0.109 &\vline&  -22.4 & 11.6 &  1  & \vline{} \vline &   Z144022-0122 & 2.244 &\vline&  -24.6 &  11.9 &  6  \\
             0244+194 & 0.176 &\vline&  -22.0 & 11.4 &  1  & \vline{} \vline &       1444+407 & 0.267 &\vline&  -22.3 &  11.5 &  1  \\
        KUV03086-0447 & 0.755 &\vline&  -23.7 & 11.9 &  8  & \vline{} \vline &    PKSJ1511-10 & 1.513 &\vline&  -23.5 &  11.6 &  2  \\
             MZZ01558 & 1.829 &\vline&  -23.0 & 11.3 &  4  & \vline{} \vline &        1512+37 & 0.371 &\vline&  -23.3 &  11.9 &  1  \\
               US3828 & 0.515 &\vline&  -22.6 & 11.6 &  8  & \vline{} \vline &     PKS1524-13 & 1.687 &\vline&  -23.9 &  11.7 &  3  \\
           Q0335-3546 & 1.841 &\vline&  -23.5 & 11.5 &  7  & \vline{} \vline &        3C323.1 & 0.266 &\vline&  -21.9 &  11.4 &  1  \\
          PKS0348-120 & 1.520 &\vline&  -24.8 & 12.1 &  2  & \vline{} \vline &       1549+203 & 0.250 &\vline&  -22.0 &  11.4 &  1  \\
           PKS0349-14 & 0.614 &\vline&  -25.2 & 12.6 & 10  & \vline{} \vline &    HS1623+7313 & 0.621 &\vline&  -22.2 &  11.4 &  8  \\
          PKS0402-362 & 1.417 &\vline&  -24.8 & 12.2 &  2  & \vline{} \vline &       1635+119 & 0.146 &\vline&  -22.2 &  11.5 &  1  \\
          PKS0403-132 & 0.571 &\vline&  -21.3 & 11.0 &  9  & \vline{} \vline &      3C345 & 0.594 &\vline&  -25.5 &  12.7 &  1  \\
          PKS0405-123 & 0.574 &\vline&  -23.1 & 11.7 &  9  & \vline{} \vline &      3C351 & 0.372 &\vline&  -24.2 &  12.2 &  1  \\
           PKS0414-06 & 0.773 &\vline&  -25.0 & 12.4 & 10  & \vline{} \vline &       1821+643 & 0.297 &\vline&  -24.9 &  12.6 &  1  \\
          PKS0420-014 & 0.915 &\vline&  -24.5 & 12.2 &  9  & \vline{} \vline &      3C422 & 0.942 &\vline&  -24.2 &  12.1 &  3  \\
           PKS0440-00 & 0.607 &\vline&  -23.3 & 11.8 &  3  & \vline{} \vline &     MC2112+172 & 0.878 &\vline&  -24.1 &  12.1 &  3  \\
            0624+6907 & 0.370 &\vline&  -24.6 & 12.4 &  1  & \vline{} \vline &     Q2125-4432 & 2.503 &\vline&  -23.0 &  11.2 &  6  \\
           PKS0710+11 & 0.768 &\vline&  -25.6 & 12.7 & 10  & \vline{} \vline &     PKS2128-12 & 0.501 &\vline&  -21.8 &  11.2 &  9  \\
        MS0824.2+0327 & 1.431 &\vline&  -23.8 & 11.8 &  7  & \vline{} \vline &     PKS2135-14 & 0.200 &\vline&  -22.9 &  11.8 &  1  \\
         MS08287+6614 & 0.610 &\vline&  -22.9 & 11.7 &  8  & \vline{} \vline &       2141+175 & 0.211 &\vline&  -22.9 &  11.8 &  1  \\
           PKS0838+13 & 0.684 &\vline&  -23.1 & 11.7 & 10  & \vline{} \vline &   Z215539-3026 & 2.593 &\vline&  -24.2 &  11.6 &  6  \\
               US1867 & 0.513 &\vline&  -25.8 & 12.8 &  1  & \vline{} \vline &       2201+315 & 0.295 &\vline&  -23.7 &  12.1 &  1  \\
             0903+169 & 0.411 &\vline&  -23.5 & 12.0 &  1  & \vline{} \vline &     PKS2204-20 & 1.923 &\vline&  -23.1 &  11.3 &  3  \\
               TON392 & 0.654 &\vline&  -23.4 & 11.8 &  8  & \vline{} \vline &   Z222702-3205 & 2.177 &\vline&  -24.7 &  11.9 &  6  \\
         MS09441+1333 & 0.131 &\vline&  -23.3 & 12.0 &  1  & \vline{} \vline &     Q2225-403A & 2.410 &\vline&  -25.0 &  12.0 &  6  \\
             0953+415 & 0.234 &\vline&  -22.3 & 11.5 &  1  & \vline{} \vline &     Q2225-403B & 0.932 &\vline&  -23.2 &  11.7 & 11  \\
             1001+291 & 0.330 &\vline&  -24.0 & 12.2 &  1  & \vline{} \vline &     PKS2227-08 & 1.562 &\vline&  -22.9 &  11.4 &  2  \\
             1004+130 & 0.240 &\vline&  -23.6 & 12.1 &  1  & \vline{} \vline &       2247+140 & 0.235 &\vline&  -23.0 &  11.8 &  1  \\
         Z101733-0203 & 1.343 &\vline&  -21.8 & 11.0 &  7  & \vline{} \vline &   Z225950-3206 & 2.225 &\vline&  -24.7 &  11.9 &  6  \\
           PKS1015-31 & 1.346 &\vline&  -24.5 & 12.1 &  7  & \vline{} \vline &   Z231751-3147 & 2.628 &\vline&  -23.4 &  11.3 &  6  \\
           PKS1018-42 & 1.280 &\vline&  -25.1 & 12.3 &  2  & \vline{} \vline &   Z232755-3154 & 2.737 &\vline&  -24.4 &  11.6 &  6  \\
             1058+110 & 0.423 &\vline&  -24.0 & 12.2 &  1  & \vline{} \vline &    PKS2345-167 & 0.576 &\vline&  -24.4 &  12.3 &  9  \\
             1100+772 & 0.315 &\vline&  -23.8 & 12.1 &  1  & \vline{} \vline &     Q2348-4012 & 1.500 &\vline&  -22.1 &  11.1 &  2  \\
\hline
\end{tabular}
\end{center}
\end{table*}

For the purposes of this work, we have to define a homogeneous compilation
of quasar host galaxy luminosities from data available in the literature.
After that, we use the rest-frame $R$-band luminosities to infer the
host galaxy stellar masses. We refer to \citet{kotilainen09} for a detailed
discussion of technicalities in the luminosity estimate of quasar host
galaxies from high-resolution imaging.

\subsection{Host galaxy luminosities from the literature}

For a complete list of data sources, we refer to the sample description
in Paper I. Apparent magnitudes in the filters of the observations are
converted to rest-frame $R$-band absolute magnitude as follows:
\begin{equation}\label{eq_mabs}
M_R = m_{\it f}-5 \log D_{\rm L}(z) - C(z) - A_{\it f}
\end{equation}
where {\it f} is the original filter of the observations, $D_{\rm L}(z)$
is the luminosity distance of the quasar in the cosmological frame we
adopted, $C(z)$ is a term accounting for filter and $k$-correction, as derived
by assuming an elliptical galaxy template \citep{mannucci01}, and $A_{\it f}$
is a term accounting for the Galactic extinction, as derived from the
\hi{} maps in \citet{schlegel98}. We remark that, in order to minimize
filter and colour corrections, we selected observations performed
using filters roughly sampling the rest-frame $R$-band. Moreover, the
$R$-band luminosity is only marginally sensitive to the age of the
stellar content. Thus, uncertainties in $C(z)$ due to the chosen host galaxy
template are negligible ($\lsim0.1$ mag) for the purposes of this work.

Low-$z$ data taken with the HST-Wide Field Camera have been analyzed
by many authors, and different $m_{\it f}$ estimates are available
for the same object and \emph{on the same data}. In particular, the studies
of \citet{bahcall97,hamilton02,dunlop03} significantly overlap onto the recent
re-analysis presented by \citet{kim08a,kim08b}.
When comparing the reported apparent host galaxy and nuclear magnitudes, we
find that the average offset is usually negligible ($\lsim0.2$ mag), but
a significant scatter is present (rms$\sim 0.3-0.5$ mag).
When more than one estimate of $m_{\it f}$ was available, we adopted the
most recent one. No images of the mid- and high-$z$ quasars in our sample were
analyzed independently by different groups, thus no superposition happens for
these objects.

\subsection{Host galaxy masses}

In order to infer the stellar mass from the host galaxy
luminosity, we have to adopt a stellar $R$-band Mass-to-Light ratio
and consider its dependence on Cosmic Time. If the majority of the
stellar population of massive galaxies did form at high redshift, as
suggested by several pieces of evidence
\citep{gavazzi02,thomas05,renzini06,cirasuolo08,cappellari09},
one may assume that the Mass-to-Light
ratio passively evolves from the formation ($z=z_{\rm burst}$)
to the present age.
On the other hand, if quasar host galaxies
suffer intense star formation episodes from $z=3$ to $z=0$ (for instance,
due to merger events), the
evolution of the stellar Mass-to-Light ratio becomes more complex
and is, in principle, different from object to
object. For the sake of simplicity, following \citet{kotilainen09}
we will consider here only the scenario of a passively evolving stellar
population with $z_{\rm burst}=5$. This is justified by the
selection of quasars with bulge-dominated host galaxies, where old
stellar populations are expected. Furthermore,
as we will discuss in section \ref{sec_discuss}, this assumption is
conservative with respect to the main results of our study.

With this caveat in mind, we find that the redshift dependence of
the host galaxy luminosity observed, e.g., in \citet{kotilainen09}
is practically removed when we take into account the evolution of the
stellar population. The stellar mass of the host galaxies
in our sample is nearly constant, with an average value of few times
$10^{11}$ \Msun{}.

Table \ref{tab_sample} lists our final estimates of the host galaxy
luminosities and masses for the quasars in our sample.

\section[]{Evolution of the \Mbh{}--\Lhost{}, \Mhost{} relations}\label{sec_results1}

\begin{figure*}
\begin{center}
\includegraphics[angle=-90, width=0.90\textwidth]{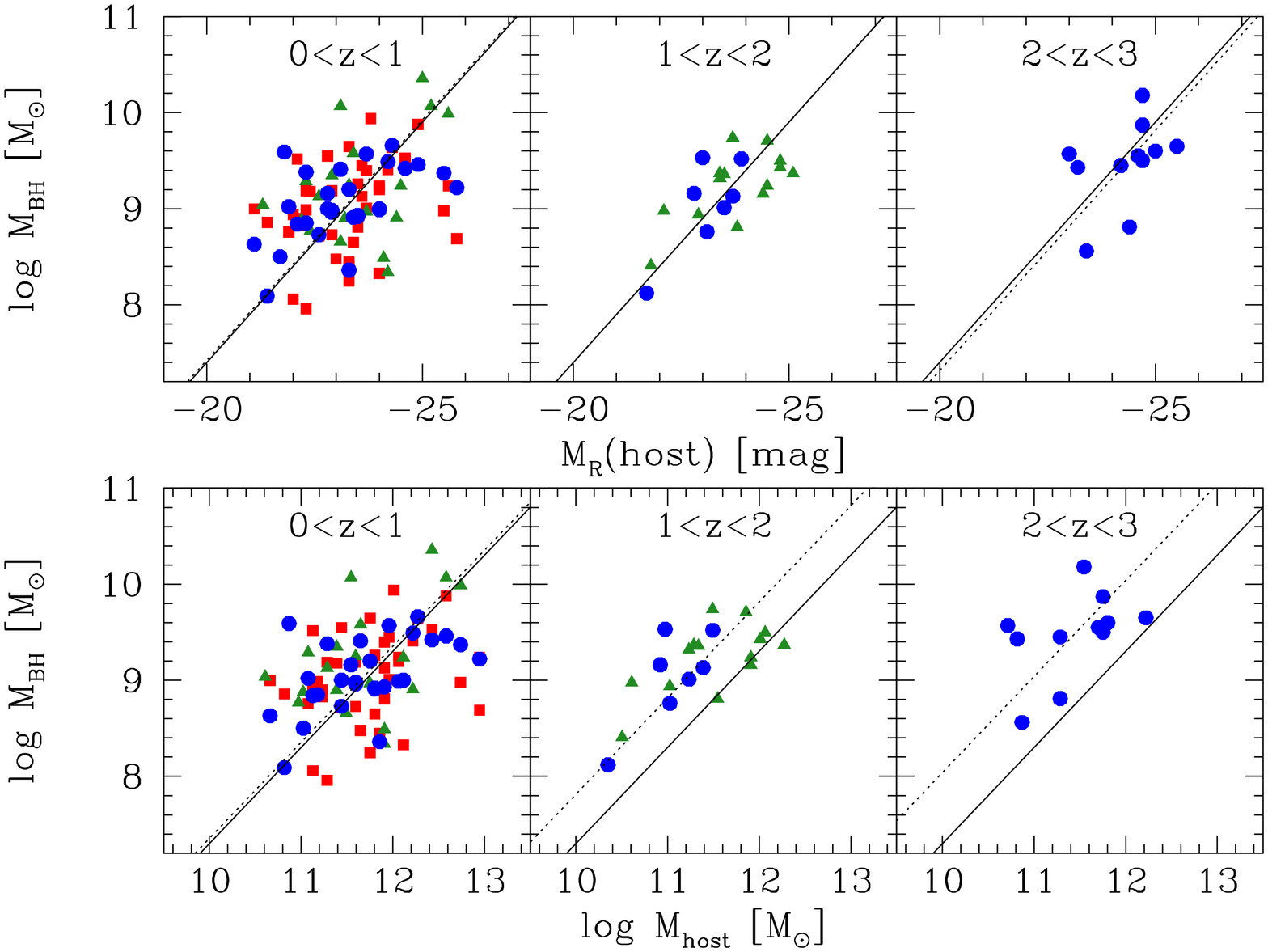}
\caption{The \Mbh{}--\Lhost{} and \Mbh{}--\Mhost{} relations
in three different redshift bins. Squares (triangles, circles)
mark quasars in which \Mbh{} is derived from \Hb{} (\Mgii{}, \Civ{}).
The reference (solid) line is the \citet{bettoni03} relation ({\it upper
panels}) or the \Mbh{}/\Mhost{}=$0.002$ case ({\it lower panels}). The dotted
line is the best fit to the data, assuming the same slope of the rest-frame
relations. No significant redshift evolution is observed when comparing \Mbh{}
with the observed host galaxy luminosities. On the other hand,
a clear offset is apparent in the \Mbh--\Mhost{} relationship
as a function of the redshift.
}\label{fig_m_m}
\end{center}
\end{figure*}

In Figure \ref{fig_m_m}, we compare our \Mbh{} estimates with
the predictions of the \citet{bettoni03} relation, defined on
$z\approx0$ galaxies, and with the expectations in
the case of a fixed \Mbh/\Mhost=$0.002$ ratio, as
observed in Local inactive galaxies \citep[see, e.g.,][]{marconi03}.

\begin{figure}
\begin{center}
\includegraphics[width=0.49\textwidth]{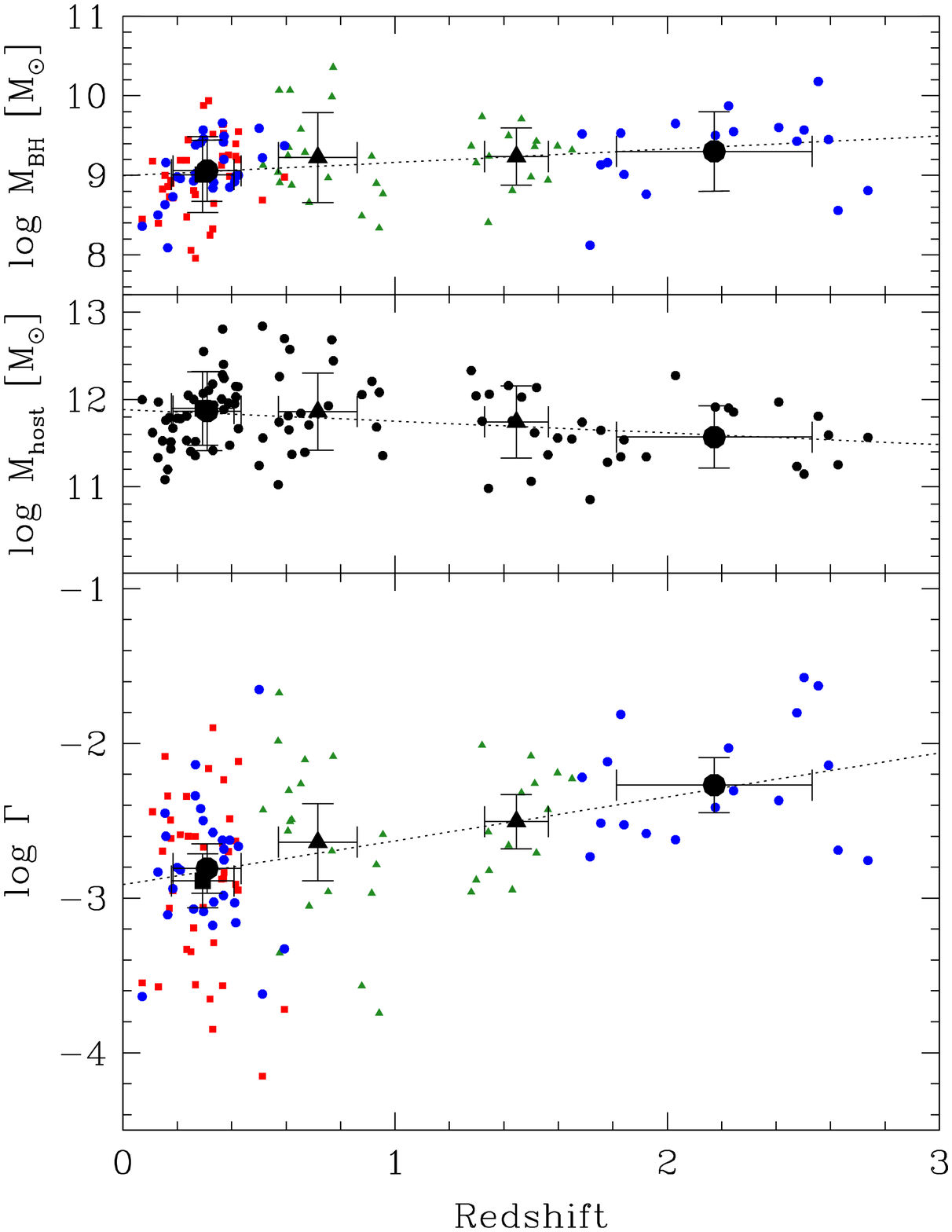}\\
\caption{The redshift dependence of \Mbh{} (\emph{top panel}),
\Mhost{} (\emph{middle panel}) and their ratio $\Gamma$
(\emph{bottom panel}). The symbol code follows Figure \ref{fig_m_m}.
The best linear fits are plotted. The average points with
rms as error bars of the \Hb{} subsample (big square), of the
low- and high-$z$ \Civ{} data (big circles) and of the \Mgii{}
data with redshift $<1$ and $>1$ (big triangles) are also shown.
}\label{fig_gamma}
\end{center}
\end{figure}

The \Mbh{}--\Lhost{} relation appears rather insensitive to the
Cosmic Time, independently on which line is adopted in the
virial estimate of \Mbh{}. When correcting for the evolution of the
stellar Mass-to-Light ratio, we find a clear increase ($\sim0.7$ dex)
of the \Mbh{}/\Mhost{} ratio with respect to what observed in the Local
Universe. In Figure \ref{fig_gamma} \Mbh{}, \Mhost{} and their ratio
$\Gamma$ are plotted all together as a function of
redshift. The linear best fit of log $\Gamma$ is:
\begin{equation}
\log \Gamma = (0.28 \pm 0.06 )\, z - (2.91 \pm 0.06)
\end{equation}
suggesting that galaxies
with similar stellar masses harbour BHs $\sim7$ times more massive at
$z=3$ than galaxies at $z=0$.

\begin{table}
\begin{center}
\caption{
Slope of the best linear fit of log $\Gamma$ = $\alpha z + \beta$, for the
whole sample and various subsamples (column 1). The number of objects in each
subsample in given in col. 2. Uncertainties on the slope (col. 3) are
analytically derived from the Least Square minimization criterion.
}\label{tab_results}
\begin{tabular}{ccccc}
   \hline
 Subset                      & \vline & N.obj. & \vline &   $\alpha$      \\
 (1)                         & \vline & (2)    & \vline &     (3)         \\
\hline
All                      & \vline & 96 & \vline & $ (0.28 \pm 0.06 )$ \\
RLQs                     & \vline & 48 & \vline & $ (0.24 \pm 0.11 )$ \\
RQQs                     & \vline & 48 & \vline & $ (0.31 \pm 0.07 )$ \\
\hline
$M_*>M_{\rm V}>M_*-1$        & \vline & 33 & \vline & $ (0.32 \pm 0.10 )$ \\
$-26>M_{\rm V}>-27$          & \vline & 31 & \vline & $ (0.32 \pm 0.11 )$ \\
\hline
N/H$>$5                      & \vline & 43 & \vline & $ (0.33 \pm 0.11 )$ \\
N/H$<$5                      & \vline & 53 & \vline & $ (0.25 \pm 0.07 )$ \\
\hline
   \end{tabular}
   \end{center}
\end{table}

\begin{figure}
\begin{center}
\includegraphics[width=0.49\textwidth]{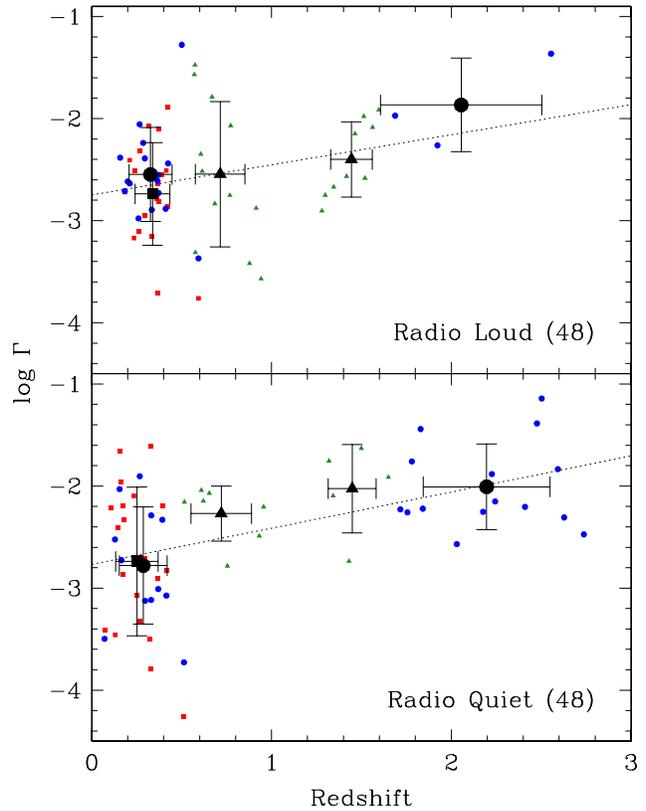}\\
\caption{The redshift dependence of $\Gamma$ for Radio Loud (\emph{top
panel}) and Radio Quiet quasars (\emph{bottom panel}) separately. The
symbol code is the same as in Figure \ref{fig_gamma}. The number of
objects in each subsample is also provided in parenthesis.
}\label{fig_gammaradio}
\end{center}
\end{figure}

\begin{figure}
\begin{center}
\includegraphics[width=0.49\textwidth]{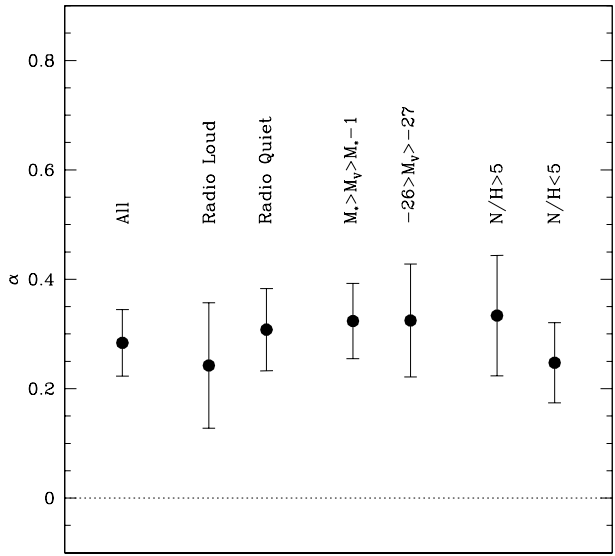}\\
\caption{The slope of the log $\Gamma$ versus $z$ linear fit in
our data, for each subsample. Error bars are the 1-$\sigma$ uncertainties
as derived from the fit algorithm (see Table \ref{tab_results}).
All the subsets have consistent slopes around $\sim0.3$ dex. A
non-evolving scenario (horizontal, dotted line) mismatches with
the observations in all the cases.
}\label{fig_valpha}
\end{center}
\end{figure}

In Figure \ref{fig_gammaradio} we study separately Radio Loud (RLQs) and
Radio Quiet quasars (RQQs),
finding that both samples are consistent with the log $\Gamma$--$z$ relation
found for the whole sample. The only remarkable difference is
in the offset, in the sense that, at any redshift, both black holes
and host galaxies in RLQs are $\sim0.2$ dex more massive than in RQQs
\citep[e.g.,][]{dunlop03,labita09c}. Table \ref{tab_results} and
Figure \ref{fig_valpha} report the slopes of the best linear fit
of log $\Gamma$ as a function of $z$ in each subsample.

\subsection[]{Is the trend of $\Gamma$ an artifact?}\label{sec_bias}

\begin{figure}
\begin{center}
\includegraphics[width=0.49\textwidth]{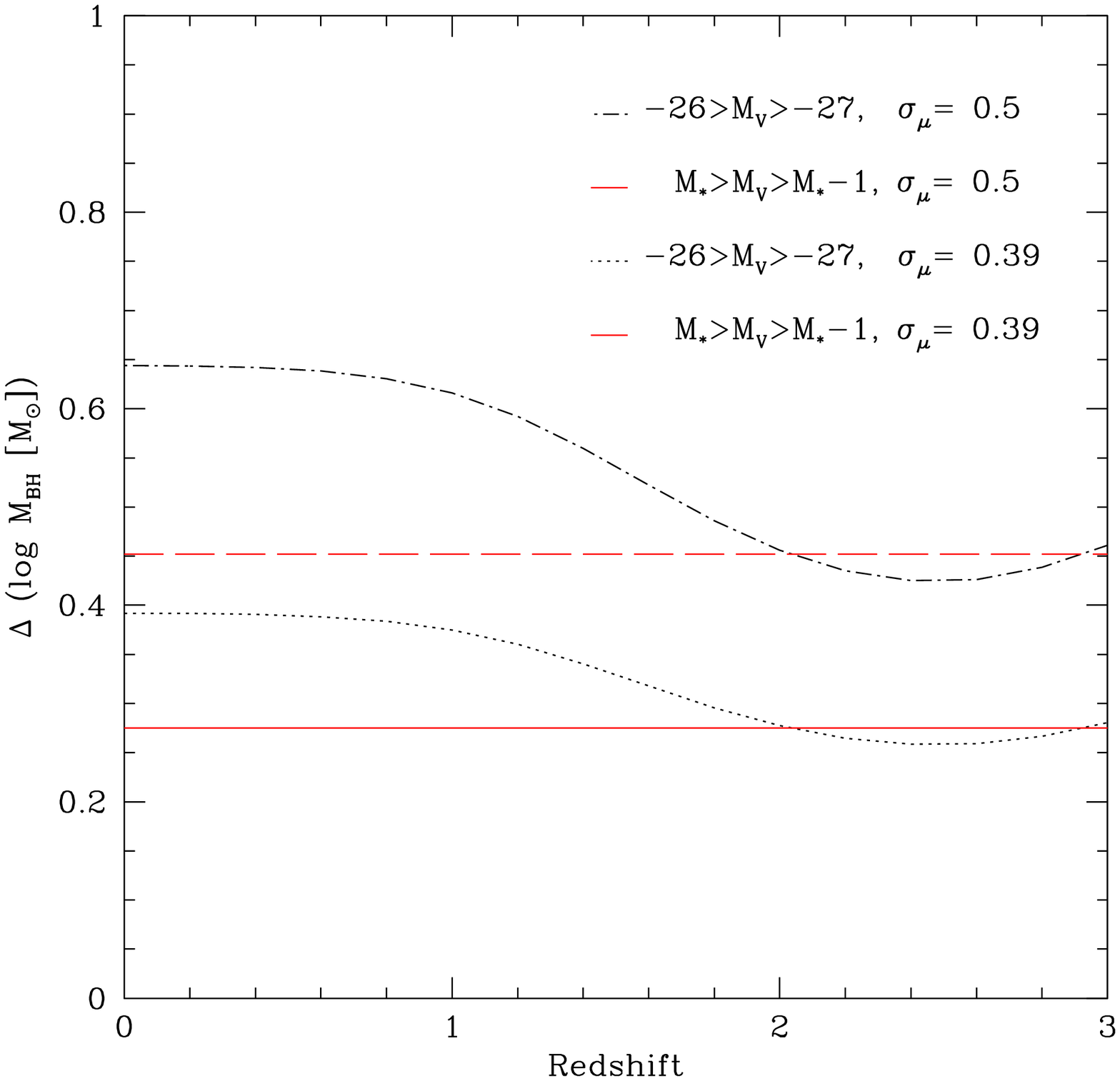}\\
\caption{The bias on the prediction of \Mbh{} from the
\Mbh{}--\Lhost{} relation with respect to the expectation from the
luminosity functions of galaxies and quasars, plotted as a function
of redshift. The bias estimates are obtained by integrating the 
luminosity function of quasars over the adopted luminosity cuts: 
$-26>M_V>-27$ (dot-dashed and dotted lines) and $M_*>M_V>M_*-1$ (dashed and
solid lines). We plot the limit cases with $\sigma_\mu=0.5$
(dot-dashed and dashed lines) and $\sigma_\mu=0.3$ (dotted and solid
lines). We note that, in the worst case, the bias increases of
$0.22$ dex (that is, a factor $1.66$) from $z=2.5$ to $z=0$.
}\label{fig_lauer}
\end{center}
\end{figure}
In this section, we discuss the possible effects that could bias
the estimate of $\Gamma$, in order to probe the reliability of the
trend observed in Figures \ref{fig_m_m}--\ref{fig_gammaradio}.

\subsubsection{The luminosity function bias}

\citet{lauer07} showed that, because of the steepness of the bright
end of the galaxy luminosity (mass) function and the presence of
intrinsic scatter in the \Mbh{}--\Lhost{} (\Mhost{}) relation, very
massive BHs are preferentially found in relatively faint (less massive)
galaxies rather than in extremely bright (massive) galaxies,
which are very rare. Since high-$z$ samples are dominated by
massive objects, the bias increases with $z$, possibly mimicking
an evolution in $\Gamma$.

In order to quantify the relevance of this bias, we assume that
\Mbh{} mostly depends on the quasar luminosity. This is
consistent with the relatively small range of Eddington ratios we
sample. If $\sigma_\mu$, the cosmic scatter of the \Mbh{}--\Lhost{},
is constant in \Lhost{}, at a given redshift the bias depends on the
shape of the luminosity function of quasars, $\Psi(M)$
\citep[see equation 25 in][]{lauer07}.
We assume the quasar luminosity function and its purely-luminosity
evolution as reported by \citet{boyle00}, basing on the 2QZ survey:
$M_*(z)=-22.0-2.5\, (1.34\, z -0.27\, z^2)$.
As a consequence, as long as we sample the same range of the quasar
luminosity function \emph{at any redshift}, the bias is kept constant.
A constant bias is irrelevant for the purposes of our study, since our
main aim is to probe the \emph{redshift dependence} of the BH--host galaxy
relations, not their absolute normalization.

\begin{figure}
\begin{center}
\includegraphics[width=0.49\textwidth]{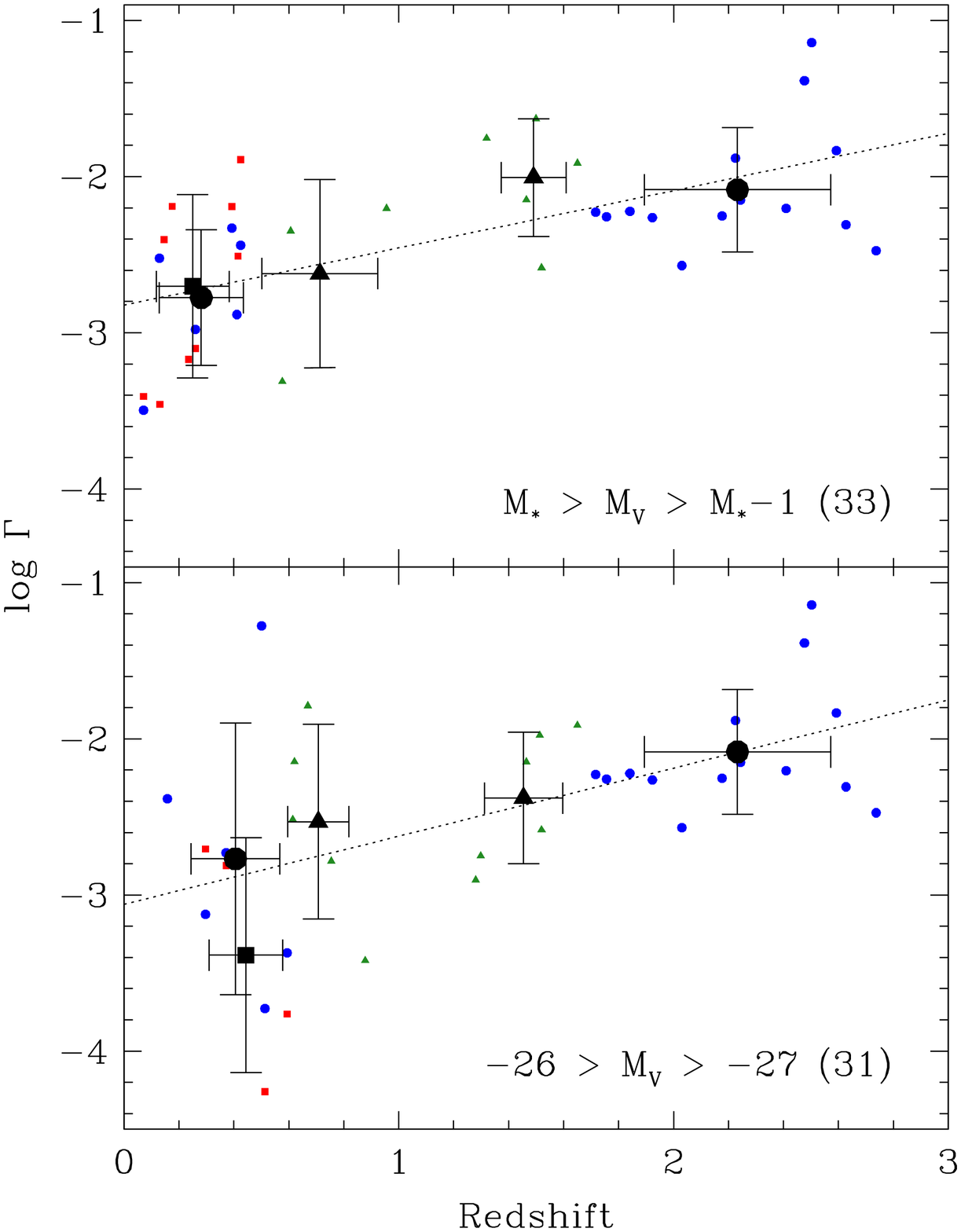}\\
\caption{The same as Figure \ref{fig_gammaradio}, for objects
with $M_*>M_{\rm V}>M_*-1$ (\emph{top panel}) and with $-26>M_{\rm V}>-27$
(\emph{bottom panel}). No significant difference is observed in the two
subsets.
}\label{fig_gammaabs}
\end{center}
\end{figure}

From Figure 1 of Paper I, we note that the constant luminosity
cut at $-26>M_V>-27$ and the $M_*>M_V>M_*-1$ cut roughly
braket the objects in our sample over 5 magnitudes in $M_V$.
Hence, if we consider the whole sample, the bias on \Mbh{} will
lie within the expectations from these two cases. Our
estimate of the redshift evolution of the bias in these two cases
is plotted in Figure \ref{fig_lauer}, for two different values of
$\sigma_\mu$, namely $0.39$ from \citet{bettoni03} and the conservative
value $\sigma_\mu=0.5$. We conclude that the bias accounts for $\lsim0.11$
dex $\approx$ a factor $1.3$ moving from $z=0$ to $z=3$. As the
observed dependence of $\Gamma$ is $\sim6$ times larger, it cannot be
explained in terms of this selection effect. As a further check, Figure
\ref{fig_gammaabs} shows the log $\Gamma$--$z$ plane only for the objects
lying in the two luminosity cuts considered in this discussion.
The observed trend is unchanged, independently on
the adopted luminosity cut (see Table \ref{tab_results} and Figure \ref{fig_valpha}).

\subsubsection{The effects of the N/H ratio}

\begin{figure}
\begin{center}
\includegraphics[width=0.49\textwidth]{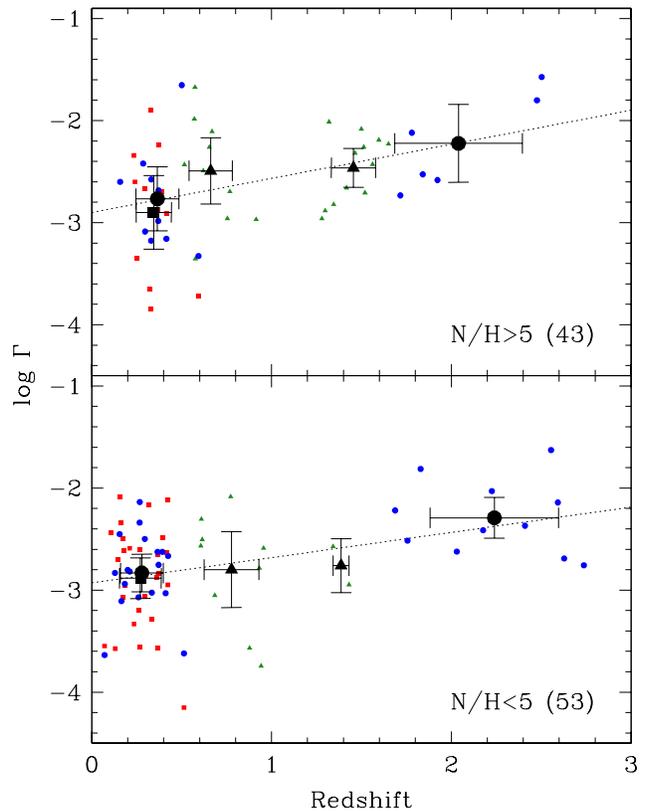}\\
\caption{The same as Figure \ref{fig_gammaradio}, for objects
with N/H$>$5 (\emph{top panel}) and N/H$<$5 (\emph{bottom panel}).
Again, no significant difference is reported.
}\label{fig_gammaN2H}
\end{center}
\end{figure}

All the objects in our reference sample are selected on the basis
of their total luminosity, which is dominated by the nuclear light in
quasars. This possibly introduces a bias in the sense that the higher
is the Nuclear-to-Host luminosity ratio (N/H), the harder is the measure
of the host galaxy luminosity, especially at high-$z$. In Figure
\ref{fig_gammaN2H} we show that the redshift dependence of $\Gamma$ in
objects with high- and low-N/H is similar. Moreover, we stress that if
we include the unresolved quasars to this analysis, the trend would be
even steeper, as they all lie at high-$z$. Analogously, this argument can be
applied for possible contaminations from disc-dominated galaxies at
high redshift, where the morphology classification may be more doubtful.
In this case, as \Mbh{} is sensitive to the bulge mass rather than to
the total galaxy mass, we should consider smaller values of \Mhost{}
for such galaxies, which would increase the value of $\Gamma$ at high $z$.

\subsubsection[]{The role of radiation pressure}

\citet{marconi08,marconi09} suggest that the virial estimates of \Mbh{} may
yield lower limits to the \emph{true} BH mass, as the radiation pressure
is not taken into account. As long as the BLR clouds are virialized,
a correction can be applied by adding a term depending
on the BLR column density $N_{\rm H}$ and the quasar luminosity.
There is still no strong constrain on the values of $N_{\rm H}$.
X-ray variability studies performed on nearby, lower luminosity
AGN suggest that the column densities may be relatively high
\citep[e.g.][and references therein]{turner09,risaliti09}, thus preventing
radiation pressure
from sustaining BLR clouds motion. However, as a clear comprehension
of the radiation pressure role in the BLR is still missing, especially
in the most luminous AGN, we limit our discussion to the following
consideration: Since the average luminosity of our data increase with
$z$, the radiation pressure effect is expected to become more severe
at higher $z$, leading to an even steeper trend of $\Gamma$ than the one
observed in Figures \ref{fig_m_m} and \ref{fig_gamma}.

\section[]{Discussion}\label{sec_discuss}

\subsection{Comparison with previous results}
\begin{figure}
\begin{center}
\includegraphics[width=0.49\textwidth]{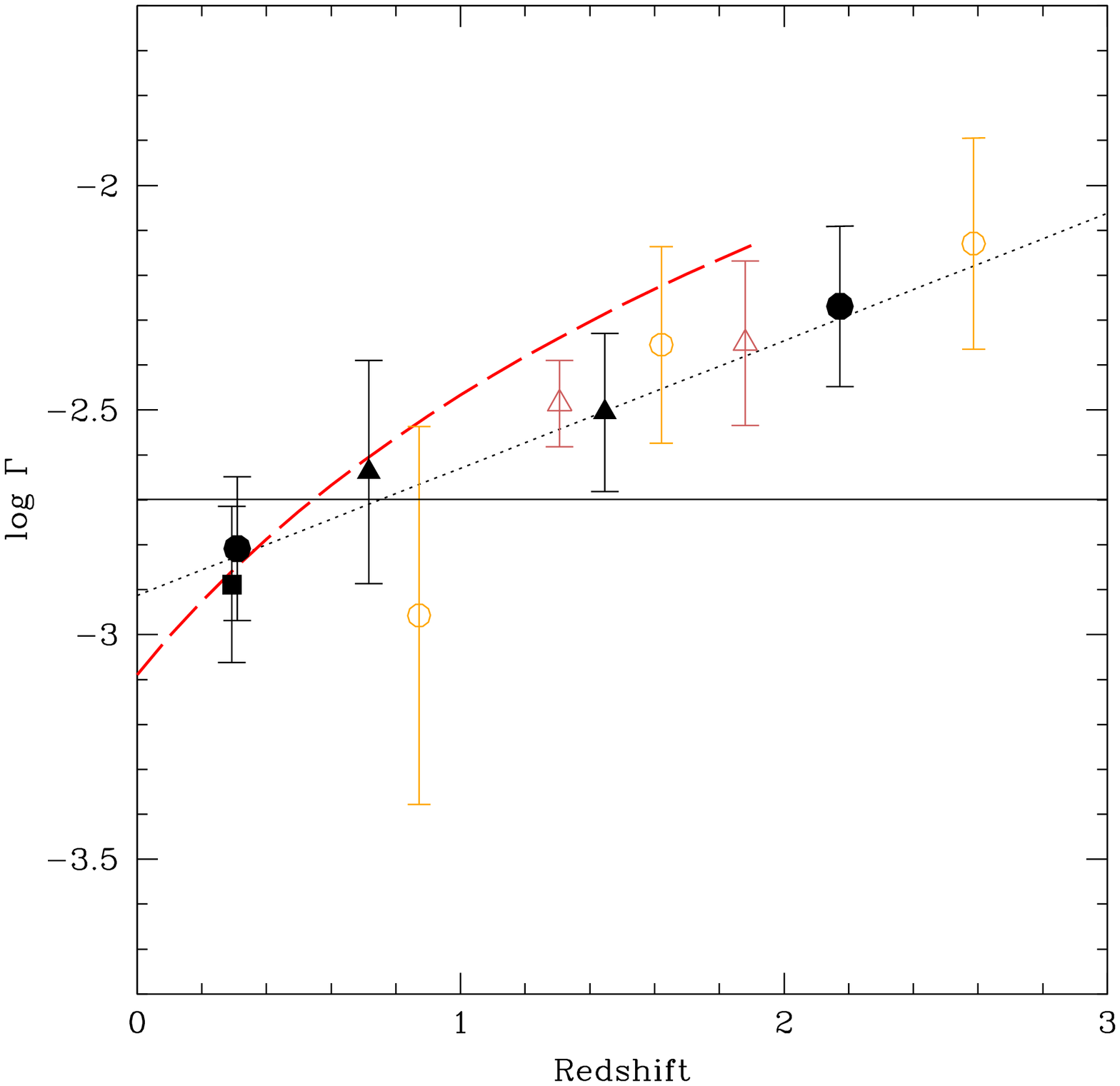}\\
\caption{The average values of $\Gamma$ for the quasars
in our sample (filled symbols), together with the linear best fit (dotted line).
Error bars are the 2-$\sigma$ uncertainties in the average values for each data bin.
For a comparison, the trend found in the study of radio loud AGN by
\citet{mclure06} is plotted as a dashed line. We also plot the average
$\Gamma$ values of the 51 lensed and non-lensed quasars from
\citet[empty circles]{peng06a,peng06b} and of the 89 AGN from
\citet[empty triangles]{merloni09}. The horizontal solid line represents
the constant $\Gamma=0.002$ case. All together the data depict a clear 
increase of $\Gamma$ with $z$.
}\label{fig_gammalit}
\end{center}
\end{figure}

As a key result of this analysis, we find that the \Mbh{}/\Mhost{} ratio
significantly increases with redshift. Hereafter we compare
these findings with those of other studies available in the literature.

\citet{mclure06} match the average trend of \Mbh{} observed in 38 Radio
Loud quasars at $z<2$ with the typical stellar masses of massive radio
galaxies in the same redshift bins. This approach relies on the assumption
that quasar host galaxies are comparable, at any redshift, with massive
radio galaxies. The major caveat here is that their results may be
biased by the different histories of quasar and radio galaxies (for instance,
note that the luminosity functions of AGN evolves differently for various
luminosity subclasses). Nevertheless, \citet{mclure06} find an increase of
$\Gamma$ comparable to the one observed in the present study (see Figure
\ref{fig_gammalit}). Our results extend these findings to
RQQs and beyond the peak age of quasar activity.

\citet{peng06a,peng06b} address the evolution of the \Mbh{}--\Mhost{}
relation as a function of redshift in $\sim20$ low-$z$ quasars and in
$\sim 30$ high-$z$ lensed quasars imaged with the HST.
Their data show a larger scatter than ours, possibly due
to uncertainties in the modelling of the lens mass distribution and the
lens light subtraction. They find that there is practically no evolution
in the \Mbh{}--\Lhost{} relation. On the other hand, when correcting for
the evolution of the stellar population, they find an excess (a factor 3--6)
in the \Mbh{} values at high-$z$ with respect to the prediction from the
local \Mbh{}--\Mhost{} relation, in qualitative agreement with our findings.

\citet{merloni09} study the \Mbh{}--\Mhost{} ratio in a sample of
89 type-1 AGN with $1<z<2.2$ from the zCOSMOS survey. Black hole masses
are derived through the standard virial assumption, while the host galaxy
luminosities and stellar masses are inferred from multi-wavelength fitting
of the spectral energy distribution of the targets (with no direct information
about the morphology of the galaxies). This technique is effective with
intermediate to low-luminosity AGN, while it cannot be applied to quasars
as bright as ours, where the nuclear light overwhelms the galaxy contribution.
They find that the average $\Gamma$ is higher than what observed in the Local
Universe, the excess scaling as $(1+z)^{0.74}$, consistently with the trend observed in our
data in the same redshift bin. \citet{jahnke09} observed 10 of the targets
in Merloni's sample with the HST and independently derived host galaxies
luminosities with a procedure similar to the one adopted in our data sources
\citep[see, e.g.,][]{kotilainen09}.
They find no evolution in the \Mbh{}--\Mhost(total) ratio. However, clues
of the occurrence of discs are present, thus the \Mbh{}--\Mhost{}(bulge) ratio
is expected to evolve as $(1+z)^{1.2}$, in agreement with our findings.

\citet{bennert09} address the \Mbh{}--\Lhost{} relation in a
sample of 23  Seyfert galaxies with $0.3<z<0.6$. They study the morphology of
the host galaxies of these objects using NIR observations from the HST. A careful
modelling is adopted in order to disentangle nuclei, bulges and disc components.
Black hole masses are derived in a way similar to that presented in paper I.
Once corrected for the evolution of the stellar population, they find
$\Gamma \propto (1+z)^{(1.4\pm0.2)}$, in good agreement with our results.

Additional indication of an evolution of $\Gamma$ comes from the
relation of \Mbh{} with the stellar velocity dispersion, $\sigma_*$, of the
host galaxy.
In particular, it is remarkable that these works suggest that the
higher is the redshift, the more massive is the black hole for a given
$\sigma_*$. For instance, \citet{salviander07} use the width of the \Oiii{}
narrow emission line as a proxy of the stellar velocity dispersion and
study the \Mbh{}--$\sigma_*$ relation in a sample of $\sim 1600$ quasars
up to $z=1.2$ taken from the SDSS. They find that \Mbh{} at high redshift
are $\sim 0.2$ dex larger than what expected from local \Mbh{}--$\sigma_*$
relation. 
A smaller evolution ($\lsim0.1$ dex), albeit with small significance,
is also proposed by \citet{shen08b}, based on a sample of 900 Type-1 AGN 
with $z\lsim0.4$.
More recently, \citet{woo08} and Woo et al. (in preparation) address the
\Mbh{}--$\sigma_*$ relation in Seyfert galaxies up to $z\sim0.6$, and find
an overall \Mbh{} excess at high redshift with respect to the prediction
from low-$z$ relationships. These findings support our results, 
notwithstanding the different characteristic luminosities, morphologies 
and stellar contents of the sampled targets with respect to those examined
in our analysis.

It is interesting to note that the $z=6.42$ quasar SDSS J1148+5251
has a black hole mass of few $\times10^{9}$ \Msun{} \citep{willott03}
and a dynamical mass of the host galaxy of $\sim5\times10^{10}$ \Msun{}
\citep{walter03,walter04}, yielding $\Gamma\sim0.1$, which is in
agreement with the extrapolation of our results at that redshift
($\Gamma\approx0.13$) and well beyond the $\Gamma=0.002$ value observed
in the Local Universe.

These results as a whole support a picture where, for a given quasar host
galaxy, its central black hole at high redshift is `over-massive' with
respect to its low-$z$ counter-parts. This picture is also consistent
with the constrains on the \Mbh{}--\Mhost{} evolution derived from the
comparison between the galaxy stellar mass function and the quasar
luminosity function \citep{somerville09}.


\subsection{Why does $\Gamma$ evolve?}

The interpretation of the observed evolution in the \Mbh{}--\Mhost{}
ratio is challenging. Exotic scenarios involving black hole
ejection from their host galaxies due to gravitational wave recoil
or to 3-body scatter may be applicable for few peculiar targets
\citep[e.g., see][but see also Bogdanovic et al., 2009, Heckman et al.,
2009 and Dotti et al., 2009 for alternative explanations]{komossa08},
but they are not applicable to the general case.
Thus, {\it if} high-$z$ quasars are destined to move towards the local
\Mbh{}--\Mhost{}, the unavoidable consequence of our results is that,
at a given \Mbh{}, galaxy masses increase from $z=3$ to the
present age. Hereafter we sketch three possible basic pictures for that.
We also present an alternative scenario, in which the fate of high-$z$
quasars may be different, the remnants of high-$z$ quasars keeping high
$\Gamma$ values down to the present age.
\begin{description}
\item[{\it Galaxy growth by mergers} --]
A first scenario involves substantial mass growth of quasar host
galaxies through merger events. It is remarkable that strong
gravitational interactions may trigger intense gas infall in the centre
of galaxies and may even lead to the activation of BH accretion.
This is observed in a number of relatively low-redshift AGN
\citep[e.g.,][]{bennert08,bennert09} showing dense close environments
or disturbed morphologies, and confirmed by the presence of young
stellar populations in some quasar host galaxies \citep{jahnke04,jahnke07}.
Two arguments disfavour this scenario. First, theoretical models based on
the structure evolution in a $\Lambda$CDM cosmology \citep[e.g.,][]{volonteri03}
predict that a massive galaxy experience only few (1--2) major merger events
from $z=3$ to $z=0$. However, our study shows that a factor $\sim7$ increase
of the stellar mass of the host galaxies is required from $z=3$ to $z=0$, which
means that the host galaxies have to suffer $\geq3$ major mergers in this redshift
range. Secondly, several pieces of evidence suggest that massive inactive
galaxies, as well as quasar host galaxies, have already formed/assembled the
majority of their mass in very remote Cosmic epochs \citep[$z\gsim3$; see, e.g.,]
[and references therein]{kotilainen09}.
The stellar population may experience episodic rejuvenation, but this
only marginally affects the mean age of the stellar content: The stellar
shells observed, e.g., by \citet{canalizo07} and \citet{bennert08} in
low-redshift quasar host galaxies
account for 5 -- 10 per cent of the total stellar population.
Similarly, in a comparison with inactive galaxies of similar mass,
\citet{jahnke04} find that quasar host galaxies are on average only
$0.3$ mag bluer. If the galaxies enter the quasar phase $\sim1$ Gyr after
the activation of the starburst, as suggested by the authors of that study,
then the involved mass is $\sim30$ per cent of the initial mass of the
galaxy. Moreover, if the quasar host galaxies contain a significant
fraction of young stellar populations, then the Mass-to-Light ratio 
would be smaller. Therefore, young host galaxies at high-$z$ would yield
a $\Gamma$--$z$ relation even steeper than that reported in Figure 
\ref{fig_gamma}.\\

\item[{\it Stellar mass growth through gas consumption} --]
Another possible interpretation is that high redshift quasar host galaxies
are gas rich, and form a significant fraction of their stellar content
in relatively recent Cosmic epochs. This is consistent with a picture in
which the black hole mass is somehow sensitive to the energetic budget
of the galaxy or its dynamical mass rather than its stellar mass
\citep[see for instance][]{hopkins07}.
This scenario is disfavoured as all the quasar host galaxies in
our sample are massive elliptical, and the stellar content of these galaxies
is usually old. Moreover, if significant star formation occurred
in quasar host galaxies in the redshift range explored in this
work, the evolution of $\Gamma$ would be much steeper, making this scenario
even less realistic.
\\

\item[{\it Evolution of the fundamental plane} --]
A number of studies suggest that inactive, massive elliptical
galaxies were more compact in the high redshift than in the Local
Universe \citep[Trujillo et al. 2006; but see also][]{cappellari09}.
In particular, an evolution of the Faber--Jackson relation is predicted,
implying that the higher is $z$, the higher is the galaxy velocity dispersion
$\sigma_*$. If \Mbh{} constantly regulates the host galaxy $\sigma_*$
\citep[e.g.,][]{silk98}, so that the \Mbh{}--$\sigma_*$ relation
does not evolve significantly, then even a small (a factor $\sim 1.6$)
increase of $\sigma_*$ for a given galaxy would yield to an excess of
a factor 7 in $\Gamma$. The main limit of this picture is that studies of
the evolution of the \Mbh{}--$\sigma_*$ relation through redshift do find
an increase of the average \Mbh{} for a given $\sigma_*$, when moving
from the Local to high-$z$ Universe \citep{salviander07,woo08,bennert09}. \\

\item[{\it Remnants of high-$z$ quasars as rare outliers} --]
We propose a scenario in which the local counter-parts of
high-$z$ quasars are high-mass outliers above the \Mbh{}--\Mhost{}
relation. The more massive is the BH, the earlier it experiences
its quasar phase \citep{marconi04,merloni04}. Our study
shows that these objects have higher expected $\Gamma$, but they 
are extremely rare, and contribute marginally to the
presently known \Mbh{}--\Mhost{} relation.
In particular, the $2<z<3$ quasars should appear as inactive massive 
galaxies with \Mbh{}$\sim10^{9.5}$ \Msun{} in the
nearby Universe. In order to quantify the occurrence of such objects
in the Local Universe, we assume the mass function of quasars:
$\Phi({\cal M}_{\rm BH}\sim10^{9.5} {\rm M_{\odot}},2<z<3)\approx4\times10^{-9}$
Gpc$^{-3}$ \Msun$^{-1}$ \citep{vestergaard09}, and take a volume 
corresponding to the most distant inactive BH for which a direct 
measurement of the mass is available\footnote{The BH at the centre 
of the brightest cluster galaxy in Abell 1836, at 147 Mpc; see 
\citet{dallabonta09}}. Under these assumptions, we expect
virtually no objects ($0.2$ in the whole volume) with such high values of
$\Gamma$.\\
Using the same arguments for targets at intermediate redshift ($1.0<z<1.5$),
we expect few ($\sim 2$) objects. Quasars at $z\sim1.2$ have $\Gamma$ values
$\sim0.3$ dex larger than that at $z=0$. This offset is close to the one
observed for the handful of objects populating the high-mass end of the local 
\Mbh{}--\Mhost{} relation \citep[e.g.][]{marconi03}.
This scenario is thus consistent both with the $\Gamma$--$z$ relation of 
quasars and with the observed shape of the local
\Mbh{}--\Mhost{} galaxy relation of nearby inactive galaxies.
\end{description}

\section[]{Conclusions}\label{sec_conclusions}

In this paper, we studied the \Mbh{}--\Mhost{} relations as
a function of redshift in a sample of 96 quasars from the
present age to $z=3$, i.e., 85 per cent of the Universe age.
We found that the \Mbh{}--\Mhost{} ratio increases by a factor
$\sim7$ from $z=0$ to $z=3$.
This trend is not affected by significant contributions due to
target selection criteria and observational biases. Moreover it is
independent of the quasar luminosity and of the radio loudness.

We interpret this trend as an indication that the most massive black
holes, living their quasar phase at high redshift, keep their high
$\Gamma$ down to the present age, becoming very rare objects
in the Local Universe. A fully consistent interpretation of these results in
terms of the common history of black holes and galaxies requires further
efforts in refining the picture sketched here. In particular,
two key points are yet to be clarified: {\it i-} how the mechanisms of quasar
feedback act onto the host galaxies; {\it ii-} which is the role of both dry
and wet mergers concerning the quasar activity and in triggering star
formation. Moreover, a better knowledge of the \Mbh{}--host galaxy relations
(improving statistics at high masses) will clarify whether very massive,
quiescent black holes can actually be found in galaxies in the Local Universe.

\section*{Acknowledgments}
We thank the anonymous referee for his/her useful suggestions which
contributed in improving the quality of the paper. We thank N. Bennert,
J.-H. Woo and A. Merloni for making us aware of their results before
the publication of their works.
This research has made use of the NASA/IPAC Extragalactic Database (NED)
which is operated by the Jet Propulsion Laboratory, California Institute of
Technology, under contract with the National Aeronautics and Space
Administration.

\label{lastpage}

\end{document}